\documentclass{aa}
\usepackage{graphicx}
\usepackage{txfonts}
\usepackage{natbib}
\usepackage{epsfig}
\begin{document}
\voffset=-1.0cm
\newcommand{\be}{\begin{equation}}
\newcommand{\ee}{\end{equation}}
\newcommand{\bea}{\begin{eqnarray}}
\newcommand{\eea}{\end{eqnarray}}
\def\d{{\rm d}}
\def\integral{{\it INTEGRAL}}
\def\rxte{{\it RXTE}}
\def\mwd{M_{\rm wd}}
\def\rwd{R_{\rm wd}}
\def\msun{M_{\odot}}
\def\ycomp{Y_{\rm Comp}}

 \title{Influence of Compton scattering on the broad-band X-ray spectra of
            intermediate polars}
\author{
V.~Suleimanov\inst{1,2},
J.~Poutanen\inst{3},
M.~Falanga\inst{4,5},
K.~Werner\inst{1}}

\offprints{V.~Suleimanov}
\mail{e-mail: suleimanov@astro.uni-tuebingen.de}

\institute{
Institute for Astronomy and Astrophysics, Kepler Center for Astro and
Particle Physics,
 Eberhard Karls University, Sand 1,
 72076 T\"ubingen, Germany
\and
Kazan State University, Kremlevskaja str., 18, Kazan 420008, Russia
        \and
            Astronomy Division, Department of Physical Sciences, 
            PO Box 3000, FIN-90014 University of Oulu, Finland
            \and
            CEA Saclay, DSM/DAPNIA/Service d'Astrophysique (CNRS FRE 2591), 91191 Gif-sur-Yvette, France
            \and 
           AIM, UMR 7158, CEA - CNRS -  Universit\'e Paris 7, France
}

\date{Received xxx / Accepted xxx}

   \authorrunning{Suleimanov et al.}
   \titlerunning{Influence of Compton scattering on X-ray spectra of IPs}

\abstract
{The majority of  cataclysmic variables  observed in the hard X-ray energy band  are intermediate polars where the magnetic field is strong enough to channel    the accreting matter to the magnetic poles of the white dwarf. 
   A shock above the stellar surface heats the gas to fairly high temperatures (10--100 keV). 
   The   post-shock region cools mostly via optically thin bremsstrahlung.} 
{ We investigate the influence of Compton
  scattering on the structure and the emergent spectrum of the post-shock region.
We also study the effect it has on the mass of the white dwarfs obtained from fitting the observed X-ray spectrum of intermediate polars.
}
{We construct the model of the post-shock region taking  Compton scattering into account. 
The radiation transfer equation is solved in the plane-parallel approximation. 
The feedback of Compton scattering on the structure  of the post-shock region is also accounted for.
A set of the post-shock region model spectra for various white dwarf masses is calculated. }
{We find that Compton scattering does not change  the emergent spectra significantly for low accretion rates or low white dwarf masses. However, it  becomes important  at high accretion rates and high
 white dwarf masses. The time-averaged, broad-band X-ray spectrum of intermediate polar V709 Cas obtained by the  \rxte\  and \integral\ observatories is fitted using the set of computed spectral models. 
We obtained  the white dwarf mass of 0.91 $\pm$ 0.02 $\msun$ and 0.88 $\pm$ 0.02 $\msun$  using models with Compton scattering taken into account and without it, respectively.  
}

{} %


\maketitle
%
%________________________________________________________________

\section{Introduction}
\label{sec:intro}

Cataclysmic variables (CVs) are close binary systems with a white dwarf (WD) as 
a primary \citep{Wr95}. A WD accretes
matter from a companion star (typically a red dwarf) that fills its Roche lobe. 
This matter forms an accretion disc inside the Roche lobe of the  WD. In 
intermediate polars  (IPs), the significant magnetic field (10$^6$--10$^7$ G) 
can  disrupt the disc at some distance from the WD surface. The matter is then 
free-falling along the magnetic field lines onto the WD and forms a strong shock 
above its surface \citep{Aizu73}. The temperature in the post-shock region (PSR) 
can be very high ($\sim$10--100 keV) and the plasma there is cooled mainly via 
optically thin bremsstrahlung radiation in the X-ray band \citep{fpr76,LM79,kl79}. 
Cooling via Compton scattering is less significant due to a relatively low 
radiation energy density in PSR. 
This contrasts with the  accretion columns of X-ray pulsars, where Comptonization is the 
dominant cooling mechanism (see \citealt{bw07}, for more details).

The theory of the PSR has been developed in a number of papers \citep{Aizu73,Wu94,WB96,Cr99}. 
The most recent investigations by \citet{can05} and \citet{Sax05,Sax07} have studied 
the role of  the two-temperature plasma and considered the dipole magnetic funneling 
(not the standard plane-parallel approximation). The plasma temperature in the PSR 
depends on the free-fall velocity at the WD surface; therefore, the X-ray spectra 
of IPs can be used to  determine the WD mass \citep{Rot81}. Such studies have been 
performed  by \citet{Cr98}, \citet{Cr99}, \citet{Ram00}, \citet{BOH00}, \citet{rev04},
\citet{sul05}, and \citet{f05}, using various X-ray data 
sets. \citet{sul05} demonstrated that,  to accurately measure the maximum 
post-shock temperature to infer the WD mass, the spectra  at high energies 
(up to at least 100 keV) are needed. 

The IPs are the most luminous and the hardest X-ray sources among accreting WDs.
 The interest in this class of objects has been growing in the past few years as IPs have
 been recently  proposed as the dominant X-ray source population detected near
 the Galactic centre by {\it Chandra} observatory \citep{muno04,ruiter06}. The IPs 
also contribute significantly to the X-ray diffuse Galactic ridge emission 
\citep{rev06}, as well as dominate the hard X-ray sky there \citep{kri07a}.
   Moreover, most of the known CVs detected by {\it INTEGRAL} \citep{bird07,kri07b} 
and {\it Swift} satellites are IPs  \citep[see e.g.][]{barlow06,bonnet07,demartino08}. 
In the strongly magnetized ($B\geq 10^{7}$ G) polar systems, cyclotron 
radiation is an important cooling mechanism,  which suppresses the high 
temperature bremsstrahlung emission, whilst it should be negligible for the IPs. 
This could explain why the majority of the  CVs observed in the hard X-ray band are IPs. 

To better understand the hard X-ray production in IPs, one needs to model their  
observed spectral properties.  The goal of the present paper is to investigate the 
influence of Compton scattering  on the structure  of the PSR and the emergent spectrum.  
We developed the physical model of the PSR and compare the computed spectral models 
to the broad-band X-ray spectrum of the intermediate polar V709 Cas  obtained 
by \rxte\ and \integral\ observatories. We have also determined the WD mass  in this system.

\begin{figure}
\begin{center}
%\centerline{
\epsfig{file=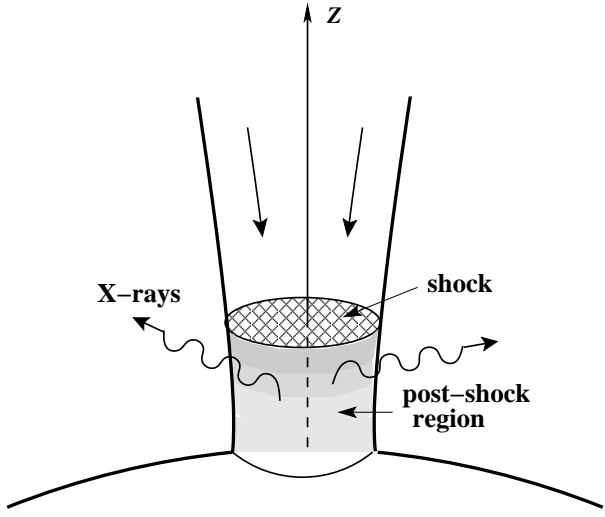,width=8.0cm} %}
\end{center}
\caption{
A schematic view of the accretion flow onto the WD magnetic polar cap. 
The accreting gas falls quasi-radially towards the magnetic poles, forming a
standing shock, below which the X-rays are emitted.}
\label{fig:geom}
\end{figure}

\section{Model of the post-shock region}

\subsection{Main equations}
\label{sec:equations}

The  structure of the stationary post-shock region  (the accretion and emission geometry is illustrated
schematically in  Fig. \ref{fig:geom}) in a plane-parallel, one-dimensional geometry can be described \citep[see e.g.][]{Cr99,sul05}  by the mass continuity equation
\begin{equation}\label{ip_u1}
\frac{\d}{\d z} (\rho V) = 0,
\end{equation}
the momentum equation
\begin{equation}\label{ip_u2}
\frac{\d}{\d z} (\rho V^2 + P) = -\frac{G\mwd}{(z+R_{\rm wd})^2} \rho,
\end{equation}
the energy equation
\begin{equation}\label{ip_u3}
  V\frac{\d P}{\d z} + \gamma P\frac{\d V}{\d z}  = -(\gamma-1)(1+A)\Lambda,
\end{equation}
and the ideal-gas law
\begin{equation}\label{ip_u4}
P = \frac{\rho kT}{\mu m_{\rm H}}.
\end{equation}
Here $m_{\rm H}$ is the hydrogen mass and $\gamma$=5/3 is the adiabatic index.
The cooling rate $\Lambda$ due to thermal optically thin radiation is given by
\begin{equation}\label{ip_u5}
\Lambda = n_{\rm e} n_{\rm i} \Lambda_{\rm N} (T),
\end{equation}
where $\Lambda_{\rm N}(T)$ is the cooling function,
$n_{\rm e}$ the electron number density, and
$n_{\rm i}$ the ion number density. We take $\Lambda_{\rm N}(T)$ for solar
chemical composition as calculated and tabulated by  \citet{SD93}.

The term $1+A$  is the amplification factor due to Comptonization. 
In contrast to the usually used integral amplification factor 
(which is just the ratio of the emergent flux to the incident flux of some seed soft photons, 
see  \citealt{is75,s84}), we use here the local amplification factor, because we are interested in 
the vertical structure of the accretion column. In a homogeneous isothermal slab, the integral and local
factors are nearly equal.  As the Compton cooling rate depends on the local plasma temperature and density 
in a different way compared to the  optically thin, thermal radiation, the local amplification factor $(1+A)$ 
is a function of the geometrical coordinate $z$.
The cooling due to cyclotron radiation is not considered here. 

Equation (\ref{ip_u1}) has the integral
\begin{equation}\label{ip_u6}
\rho V = a, 
\end{equation}
where the mass accretion rate $a$  per unit area is a free parameter of the model.

A simple analytical solution of Eqs. (\ref{ip_u1})--(\ref{ip_u4})
 can be obtained \citep[see][]{FKR}, if one assumes a  constant gas pressure throughout the PSR 
and  considers the cooling only by bremsstrahlung. 
The distributions of the velocity, temperature, and density along the PSR height $z$ 
in this case are 
\be \label{ip_ua31}
V  = V_0 \left
( \frac {z}{H_{\rm s}} \right )^{2/5},~~~~
T  =  T_0 \left ( \frac {z}{H_{\rm s}} \right )^{2/5},~~~~
\rho =  \rho_0 \left ( \frac {z}{H_{\rm s}} \right )^{-2/5},
\ee
where the shock height above the surface, 
\be \label{ip_ua30}
H_{\rm s} = 3.94 \times 10^7\ a^{-1}\ \mu^{-1/2}\  \mu_{\rm e}^2 m^{3/2} R_9^{-3/2} ~~\mbox{cm}, \\ 
\ee
and the parameters at the shock position are
\bea \label{ip_ua32}
  V_0 & = & 1.29 \times 10^8~m^{1/2}~R_9^{-1/2}~~ \mbox{cm\ s}^{-1} , \nonumber \\
 \rho_0 & = & 7.75 \times 10^{-9}~a~m^{-1/2}~R_9^{1/2}~~ \mbox{g\ cm}^{-3}, \\
T_0 & = & 6.03 \times 10^8~\mu~m~R_9^{-1}~~ \mbox{K}. \nonumber
\eea
Here   $\mu$ is the gas mean molecular
weight (in units of $m_{\rm H}$)
and $\mu_{\rm e}$   is the number of nucleons per electron in the fully ionized plasma 
($\mu=0.62$ and $ \mu_{\rm e}=1.2$ for solar composition). We use the dimensionless variables for the WD mass $m = \mwd/\msun$ and the radius $R_9 = \rwd/10^9$cm. The WD radius can be calculated from  the  WD mass-radius relation \citep{Nau72}:
\be  \label{eq:rwd}
R_9  = 0.78     \left [ \left( \frac {1.44}{m}\right)^{2/3} -   \left( \frac{m}{1.44} \right )^{2/3}  \right ]^{1/2}.
\ee
Alternatively, for masses   $m$ in the range 0.5--1.2, one can use a linear approximation to the mass-radius relation
\be \label{eq:rwd_linear}
R_9 = 1.364-0.807m.
\ee

\subsection{Numerical scheme}
\label{sec:method}

We now describe the method of numerically solving the PSR structure.  First of all, 
the input parameters, the WD mass, and the local accretion rate $a$ are defined. Then 
we use two-loop iteration scheme. In the outer loop the amplification factor is changed, 
while in the inner loop the
PSR structure and the emergent spectrum are computed with the fixed
distribution of the amplification factor on the relative height $z/z_0$, where $z_0$ is the shock position.
First, an initial model without Comptonization ($A(z)=0$) is
calculated.  Equations (\ref{ip_u1})--(\ref{ip_u4}) are solved by the
shooting method from $z=z_0$ to $z=0$ with the following boundary
conditions at $z=z_0$ (see details in \citealt{sul05}):
\bea
\label{ip_u10}
   V_0  & = & 0.25 \sqrt{2G\mwd/(\rwd+z_0)}, \nonumber \\ 
   \rho_0 & = & \frac {a}{V_0}, \\ \nonumber
   P_0 & = & 3 a V_0,      \\ \nonumber
T_0 & =  &3 \frac {\mu m_{\rm H}}{k}  V_0^2.
\eea
The shock position $z_0$ is found iteratively from the additional boundary condition: $V=0$ at the WD surface. The obtained PSR structure along $z$-coordinate is then considered as a
plane-parallel atmosphere. 
The radiative transfer equation in this atmosphere is solved at a grid of 90 column densities
$\Sigma$ (defined in usual way $\d \Sigma = -\rho \d z$), distributed equidistantly  from  $\Sigma\approx
 10^{-6}$ g cm$^{-2}$ to $\Sigma_{\rm max} = \Sigma(R_{\rm WD})$.
The Compton amplification factor $A$ as a function of the PSR height 
is given by the ratio of the radiation energy loss rate, obtained from the solution of the 
radiative transfer equation, to the one given by the thermal cooling function:
\be
\label{ipc_u10}
  1+A(z) = \frac{\d F/\d z}{\Lambda} = - \rho\frac{\d F/\d \Sigma}{\Lambda}, 
\ee
where $F$ is the integrated over frequency radiation flux.

Then we again solve Eqs. (\ref{ip_u1})--(\ref{ip_u4}) with the computed amplification 
factor. Here we assume that the amplification factor only depends on the relative position 
$z/z_0$ in the PSR. It is important, because in the inner loop of iterations the 
shock position and the set of geometrical depth $z$ change from iteration to iteration.  
Therefore, for the given shock position $z_0$, the amplification factor at height $z$ is 
 computed by interpolating the dependence $A(z/z_0)$. This iteration procedure is repeated 
until the relative change in the shock position is less than 1 per cent (with 5--6 
iterations being sufficient). As a result we obtain the self-consistent PSR model with 
Compton cooling, together with the emergent spectrum of radiation.

\subsection{Radiative transfer with Compton scattering}

The radiative transfer equation for specific intensity $I(x,\mu)$ with
Compton scattering taken into account in the plane-parallel approximation
has the form
\be
\label{ipc_u1}
  \eta \frac{\d I(x,\eta)}{\d \Sigma} = [\sigma(x)+k(x) ][ I(x, \eta) -
S(x, \eta)] ,
\ee
where  $\sigma(x)$ and $k(x)$  are electron scattering and true absorption 
opacities at depth $\Sigma$,  $\eta$ is the cosine of the angle between the 
normal to WD surface and the direction of the radiation propagation. 
The absorption opacity is the sum of the free-free opacities of the fully 
ionized 25 most abundant chemical elements with solar composition.
The source function is the sum of the thermal part and the scattering part:
\bea
\label{ipc_u2}
& &S(x, \eta)  =  \frac{k(x)}{\sigma(x)+k(x)}B(x) +
\\ \nonumber & & \frac{x^2}{\sigma(x)+k(x)} \int_0^\infty
\frac{\d x_1}{x^2_1} \int \d\varphi \int  \d \eta_1 
R(x,x_1, \psi) I(x_1, \eta_1),
\eea
where $B(x)$ is the Planck function.
The Compton scattering redistribution function that describes the electron scattering 
(in the isotropic Thomson approximation in the electron rest frame, which is 
 accurate enough for electron temperatures below 100 keV), is \citep{an80,p94,ps96}:
\be
\label{ipc_u3}
    R(x,x_1,\psi) = \frac{1}{8\pi Q}
\frac{e^{-\gamma_*/\Theta}}{K_2[1/\Theta]},
\ee
where
\be
\label{ipc_u7}
x_1=\frac{h\nu_1}{m_{\rm e} c^2}, \quad   x = \frac{h\nu}{m_{\rm e} c^2} 
\ee
are dimensionless photon energies before and after scattering,
\be
 \label{ipc_u8}
 \Theta = \frac{kT}{m_{\rm e}c^2}
\ee
is the electron temperature in units of the electron rest mass, 
 $\psi= \eta \eta_1 + \sqrt{1-\eta^2}\sqrt{1-\eta^2_1}\cos\varphi$ is the cosine of the scattering angle,
  $K_2$ is the modified Bessel function,   and
\be
\label{ipc_u4}
         \gamma_* = \frac{Q}{\sqrt{2xx_1(1-\psi)}}, \quad 
      Q^2 = x^2+x_1^2 - 2xx_1 \psi . 
\ee

The outer boundary condition is found from the lack of incoming radiation at the PSR surface, and the inner boundary condition is taken as the Planck function with the effective temperature of the WD ($\approx 2 \times 10^4$ K).
The formal solution of the radiation transfer equation (\ref{ipc_u1}) is obtained 
using the short-characteristic method \citep{ok87}, and the full solution 
is found with a simple $\Lambda$-iteration method. This method gives a fast 
solution in this case, as the PSR has a rather small electron 
scattering optical depth ($\tau_{\rm e} \le 1$). This is also  why 
we do not need to consider the nonlinear terms (i.e. induced scattering) 
in the radiation transfer equation.

\begin{figure}
\centerline{\epsfig{file=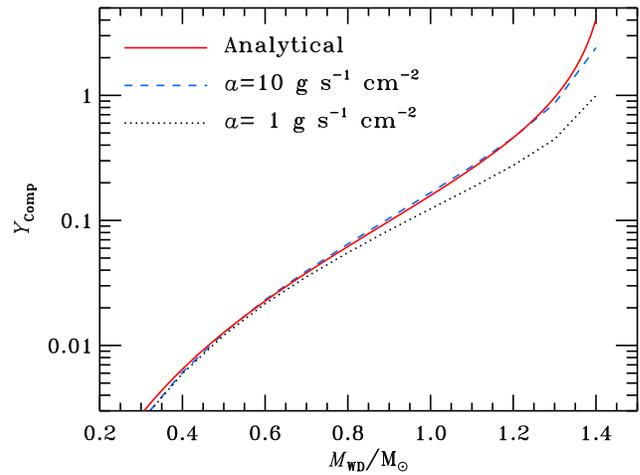,width=8.5cm}}
\caption{
Dependence of the Compton parameter $\ycomp$ for the PSR 
on the WD mass.  The solid curve gives the results for the analytical model  (\ref{ip_ua31}) of \citet{FKR}  
given by Eq. (\ref{ip_ua33}), while 
the dotted and dashed curves are for the numerical model of \citet{sul05} 
for two accretion rates $a$ = 1 and 10 g s$^{-1}$ cm$^{-2}$, respectively. } 
%\end{center}
\label{fig:ycomp}
\end{figure}

\begin{figure*}
%\centerline
\begin{center}
\leavevmode \epsfysize=6.0cm \epsfbox{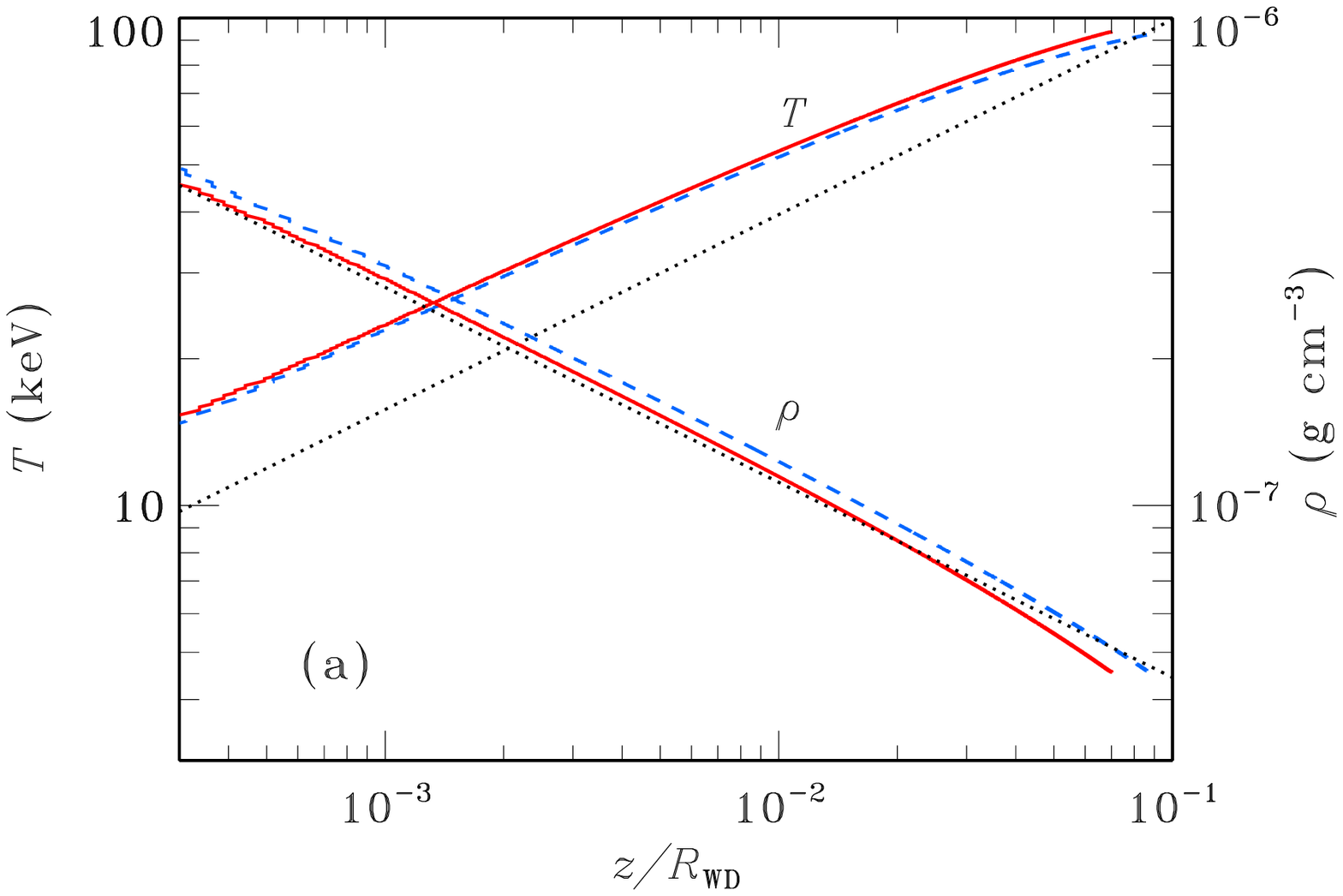} \hspace{1cm}
\epsfysize=6.0cm \epsfbox{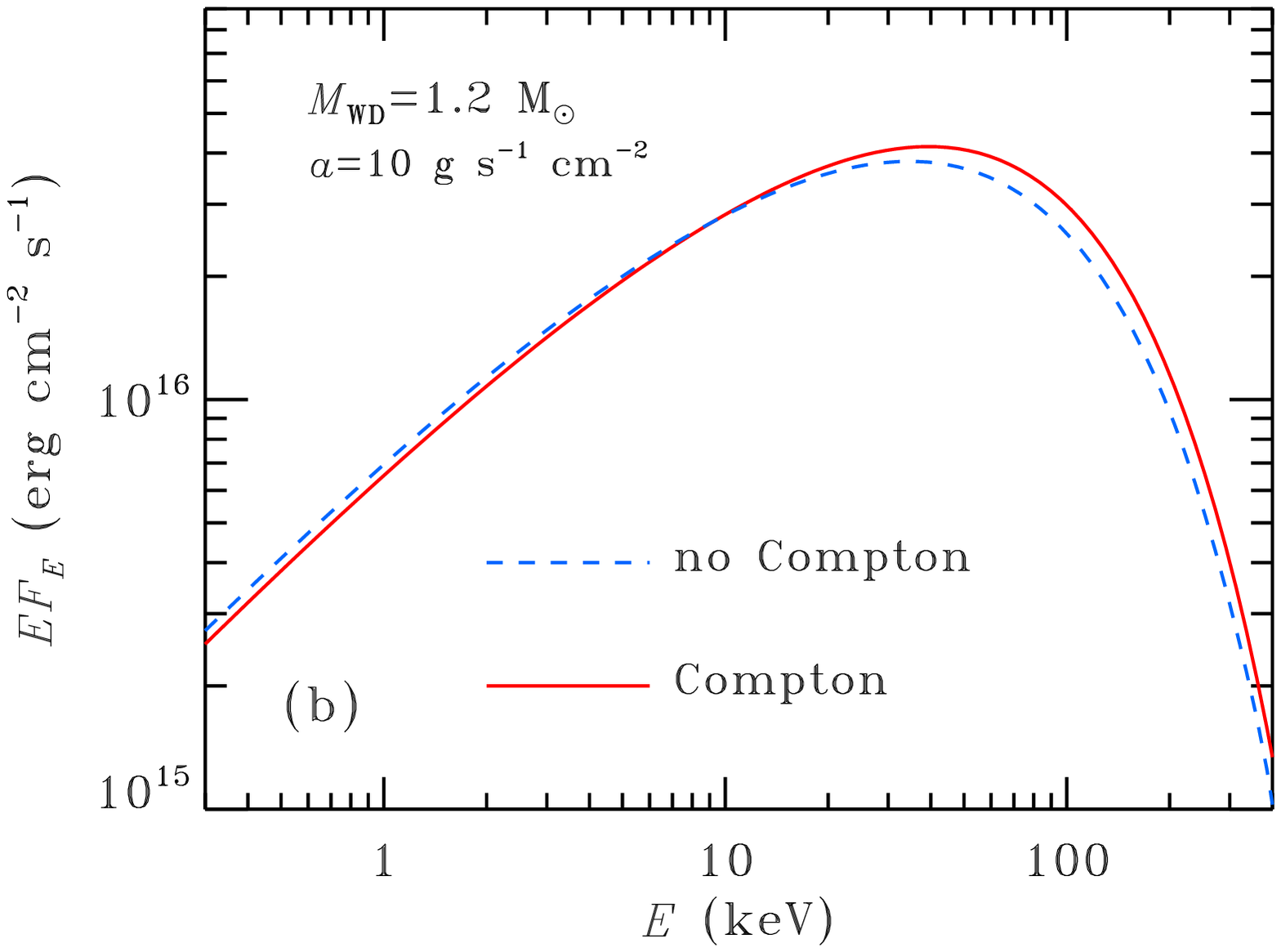}
\end{center}
\caption{
Comparison of PSR models computed for $\mwd=1.2 \msun$ with a  local accretion rate
$a = 10 $ g s$^{-1}$ cm$^{-2}$ with (solid curves) and without
(dashed curves) Compton scattering taken into account. The
temperature and the density structures (panel a), as well as the emergent spectra
(panel b), are shown. The dotted lines correspond to the analytical solution (\ref{ip_ua31}).
} 
%\end{center} 
\label{fig:psr}
\end{figure*}

\subsection{Results of calculations}
\label{sec:results}

Before presenting the results of our calculations, let us first evaluate the importance of   cooling  by Compton scattering in the PSR.  The Compton parameter 
\be \label{ipc_u13}
  \ycomp =  4 \Theta \max(\tau_{\rm e}, \tau_{\rm e}^2) 
\ee
can be calculated for various models of the PSR. 
The analytical model  (\ref{ip_ua31}) yields
\be \label{ip_ua33}
  \ycomp = 4 \sigma_{\rm T}  \int_{0}^{H_{\rm s}}~ \Theta ~n_{\rm e}~ \d z =  
  0.05~\mu^{1/2}~\mu_{\rm e}~m^2~R_9^{-2} .
\ee
It is clear the Compton parameter can be close to unity  only for high-mass
WDs.  It is possible to also compute $\ycomp$ using  numerical
models of the PSRs \citep{sul05}. The dependence of $\ycomp$ on the WD mass is shown in
Fig.~\ref{fig:ycomp}. We see that at high accretion rates and high WD masses, Compton 
scattering can affect  the structure of the PSR and the emergent spectrum.
 
The influence of Comptonization on the temperature and the density structure of the PSR, 
as well as the emergent spectra, are demonstrated in Fig.~\ref{fig:psr}. Here we take  a 
1.2 $\msun$ WD with a  local accretion rate $a = 10 $ g s$^{-1}$ cm$^{-2}$ and compute 
the PSR structure with and without Comptonization taken into account. The amplification 
factor $1+A$ due to Compton scattering is about 1.5 at the top  of the PSR (near the shock) 
and decreases to 1 at the WD surface. Therefore,  if Comptonization is taken into account, 
the cooling rate becomes higher, the height of PSR  smaller, and the emergent spectrum 
slightly harder. But the difference between the emergent spectra with and without 
Comptonization is, however,  rather small.

We computed the grid of PSR models for WD masses ranging from 0.3 
to 1.3 $\msun$ with step 0.02 $\msun$ (51
models in total).  The total accretion rate was assumed to be $\dot M = 10^{17}$ g s$^{-1}$ 
with the fixed part $f = 10^{-3}$ of the WD surface occupied by PSR.  This 
gives different local accretion rates $a$ for various WD masses, varying from 
$\sim 0.5$ to $\sim 5$ g s$^{-1}$ cm$^{-2}$. The emergent spectra of PSR with 
different local accretion rates (in the interval 0.1--10 g s$^{-1}$ cm$^{-2}$) 
are very similar \citep{sul05}; therefore, we have only one free parameter, the WD 
mass, in our grid of model spectra.  This means that the mass can be found 
by fitting the observed X-ray spectrum.  Of course, the WD mass determination
using spectral fitting is model dependent, because of using of the specific 
theoretical WD mass-radius relation and of the framework the theoretical spectra are calculated. 
However, the WD mass determination depends very little on the particular choice  of $\dot M$ and  $f$.

 The grid of the IP spectral models has
been incorporated into the {\sc XSPEC} software package  under the  name {\sc IPCOMP} 
and is publicly available.

\section{X-ray spectrum of  V709 Cas}
\label{sec:application}

V709 Cas is  a typical IP emitting in the hard X-rays \citep[see e.g.][]{sul05}. The 
present dataset was obtained using the \rxte/PCA (3--20 keV) \citep{jahoda96} and 
\integral/IBIS/ISGRI (20--120 keV) \citep{u03,lebr03} instruments with the total 
effective exposure times of 47 ks and 4.5 Ms, respectively.  Our spectral model 
consists of the PSR model {\sc IPCOMP}, an iron line modelled as a Gaussian, and 
the partial absorption model {\sc PCFABS} described by the covering factor 
$C_{\rm F}$ and the column density $N_{\rm H}$. The spectral analysis is performed 
with the {\sc XSPEC} \citep{Arnaud96} version 12.4.

\begin{figure}
\centerline{\epsfig{file=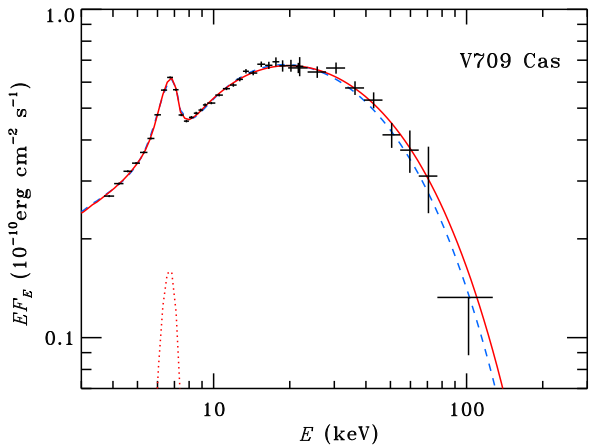,width=8.0cm}}
\caption{The unfolded spectrum of V709 Cas and the best-fit spectral model of the PSR. 
The crosses represent the data from  \rxte/PCA (3--20 keV) and \integral/ISGRI (20-100 keV) instruments. The total spectrum of the PSR model  is shown by the solid (with Compton scattering) and the dashed (without Compton scattering)  curves. The dotted curve represents the Gaussian iron line.
} 
\label{fig:spectrum}
\end{figure}

The best fit  with $\chi^{2}/{\rm d.o.f.} = 31/30$ is obtained for $\mwd=0.91_{-0.02}^{+0.03} \msun$,  
a partial covering fraction  $C_{\rm F}=0.41\pm0.05$ and $N_{\rm H}=(53\pm10) 
\times 10^{22}$ cm$^{-2}$ (see Fig. \ref{fig:spectrum}).  The iron line energy 
is $6.5\pm0.2$ keV with  an equivalent width of 508 eV. The  unabsorbed flux 
between 0.5 and 150 keV is $F \simeq 2.4\times10^{-10}$ erg cm$^{-2}$ s$^{-1}$, 
which translates into  a bolometric luminosity of  $L \simeq 1.52\times10^{33}$ 
erg s$^{-1}$, for the source distance of 230 pc \citep{bb01}. 

We can estimate the mass  accretion rate using  the simple relation $\dot M = L \rwd/(G\mwd)$. 
We get $\dot M=0.8\times10^{16}$ g s$^{-1}$ for the WD radius of $\rwd\simeq0.61\times10^{9}$ cm, 
which was computed using Eq. (\ref{eq:rwd}). To investigate the role of  Compton scattering, we 
also implemented  into {\sc XSPEC} the grids of the PSR model without Compton scattering. 
The derived parameters are similar to those obtained with the  Compton scattering model with a 
comparable reduced $\chi^{2}$.  The WD mass is found to be slightly lower $\mwd=0.88_{-0.01}^{+0.02}\msun$. 
This seems counter-intuitive, because Comptonization increases the flux at high energies, and therefore 
would need lower temperature (and therefore lower WD mass) to radiate the same energy. However, 
the spectral slope in the 4--20 keV energy range is different for the two models, so the 
absorber parameters also change to $C_{\rm F}=0.44\pm0.05$ and $N_{\rm H}=(63\pm10) \times 10^{22}$ cm$^{-2}$. 
Thus for V709 Cas, both mass estimations are consistent with each other within the errors, and 
Compton scattering does not have much effect, which is expected for such a WD mass and a rather 
low accretion rate.  
 
For illustrative purpose, we also obtained  a fit of the observed spectrum with a single temperature 
bremsstrahlung model. We obtained $T= 29.6 \pm 2.5$ keV, $C_{\rm F}=0.31\pm0.04$,  $N_{\rm H}=(58\pm15)
 \times 10^{22}$ cm$^{-2}$, and the iron line energy $6.5\pm0.2$ keV with the reduced $\chi^2$ = 1.062. 
Using a linear approximation to the mass-radius relation (\ref{eq:rwd_linear}), we can estimate  the WD mass: 
\be \label{uadd_1}
    m = \frac{T_{\rm keV}}{23.6+0.59T_{\rm keV}},
\ee
which gives $\mwd = 0.72 \pm 0.035 \msun$.
This result illustrates the error from the use of a simple one-temperature bremsstrahlung model for 
the WD mass estimation.

\section{Conclusions}
\label{sec:conclusions}

We have studied the influence of Compton scattering on the structure of the post-shock 
region and the broad-band X-ray spectra of intermediate polars.  Compton scattering  
leads to slightly harder emergent spectra and smaller PSR height. The effect can be 
significant for the luminous intermediate polars with high-mass WDs. We have also 
constructed a grid of the emergent spectra 
in a wide range of WD masses and incorporated into the {\sc XSPEC} package.  We 
used this grid to fit  the broad-band X-ray spectrum of V709 Cas observed by \rxte\  
and \integral\ observatories and  determined the WD mass in this object. 
Accounting for Compton scattering does not change the obtained mass significantly. 

In this work we have not considered the cooling by cyclotron radiation. 
However, for WDs with stronger magnetic field (polars), Comptonization of the cyclotron 
photons can significantly affect  the structure  of the PSR and the emergent spectra.  
We plan to perform such a  study in the future.

\begin{acknowledgements}  
VS was suppoted by DFG (grant We 1312/35-1 and grant SFB / Transregio 7 ``Gravitational Wave Astronomy'')
 and by the President's programme  for support of leading
science schools (grant NSh--4224.2008.2). 
JP has been supported by the Academy of Finland grant 110792. 
MF acknowledges the French Space Agency (CNES) for financial support
and  thanks  J. M. Bonnet-Bidaud for valuable discussions.
VS, JP, and MF acknowledge the support of the International Space Science
Institute (Bern), where part of this investigation was carried out. 
\end{acknowledgements}

\bibliographystyle{aa}

\end{document}